\documentclass[12pt,english]{article}
\usepackage{newcent}
\usepackage[T1]{fontenc}
\usepackage[latin9]{inputenc}
\usepackage[letterpaper]{geometry}
\usepackage{amsthm}
\usepackage{amsmath}
\usepackage{amssymb}
\usepackage{graphicx}
\usepackage{setspace}
\usepackage{esint}

\makeatletter
\numberwithin{equation}{section}
\numberwithin{figure}{section}

\makeatother

\usepackage{babel}
\begin{document}
\begin{doublespace}

\title{An Introduction to Quantum Computing.}
\end{doublespace}

\begin{doublespace}

\author{Zachary Burell}
\end{doublespace}

\maketitle
\begin{doublespace}

\section{Introduction. }
\end{doublespace}

\begin{doublespace}
Physics has often progressed very rapidly as the precision of measurements
has increased. For instance, it was the precise measurements of Tycho
Brahe which were instrumental in Kepler's deduction of the elliptic
orbit, a result which later formed a cornerstone of Newton's universal
law of gravitation. There is a minimum level of precision in the measurement
of the planetary orbits, below which it becomes impossible to distinguish
between elliptical orbits with minuscule eccentricity, and circular
orbits. Once data of the requisite precision to notice a difference
was available, all that was left for someone to do is to put the pieces
together, and \emph{voila} you get Newton!

Quantum Optics, plays an essential role in quantum metrology, a field
in which the level of precision has increased exponentially over the
past two decades. New techniques of increasing precision in quantum
optics have increased the significant digits of some of the experimentally
measured fundamental constants by orders of magnitude. Quantum optics
is so robust that it also of immediate use in testing theories of
gravity and quantum field theory, for example, the L.I.G.O. collaboration
are using a 2 km baseline Michelson Interferometer to search for gravitational
radiation.

These unheralded successes are currently pushing into new domains
of experimental precision, and we now have more direct access to the
deeper layers of nature. Every week new quantum computing components
are brought into being, by the shear effort of those working in the
field. With each new switch, isolation mechanism, algorithm, etc.,
the goal of scalable robust quantum computing becomes more eminent.
If we are successful in constructing quantum computers, the effect
will be more revolutionary than anything before, including the classical
computer and the internet. The vast expanse of Hilbert Space will
then be in the throes of man.

The fields of quantum optics and quantum computing are closely related
to one another. Very often breakthroughs in quantum optics are implemented
in quantum information processing, storage and quantum communication
devices. For example, two ways in which cavity QED techniques may
be used to to perform quantum computations are (from {[}8{]}) 
\end{doublespace}
\begin{enumerate}
\begin{doublespace}
\item Quantum information can be represented by photon states, with atoms
trapped in cavities providing the non-linear interactions between
photons, necessary for entanglement. 
\item Quantum information can be represented by atoms in different states,
where photons are used to communicate between the different atoms/states.\end{doublespace}

\end{enumerate}
\begin{doublespace}
Any realization of these schemes would at some point have to address
the problem of precision control of population transfer, as a means
to generate single photons. Such precision is a per-requisite for
realizing any completely quantum technology, that is, any technology
based on computational components whose functionality depends \emph{a'priori}
on quantum non-linearities, an example of which is entanglement.
\end{doublespace}

\begin{doublespace}

\section{Field Quantization.}
\end{doublespace}

\begin{doublespace}
This treatment of the field quantization will closely (but not exactly)
follow chapter 2 of Gerry et.al., given in {[}1{]}. In order that
we understand the interaction of quantized modes of the electromagnetic
field with ``$atoms$'', (whose definition will, for the moment
remain general; we will define an atom to be any bound sate of electrons
in a potential $V\left(\mathrm{r}\right)$.) we must first understand
the properties of the quantized fields themselves. In the following
we begin with the simple case of a single mode field confined to a
1-d cavity. This clearly represents an idealized situation, but we
will later generalize to the case of a multimode field in some three
dimensional cavity.
\end{doublespace}

\begin{doublespace}

\subsection{Single mode field. }
\end{doublespace}

\begin{doublespace}
We begin as always, with the one-dimensional square well, but in the
context of quantized modes of the electromagnetic field, which will
be relevant for our later analysis of quantized modes of optical cavities
, etc. One fruitful and interesting scenario to investigate for our
purposes, is the case of a radiation field confined to a one dimensional
cavity free of sources(i.e. there are no currents,charges, or any
dielectric media in the cavity), oriented along what we choose to
be the $z$ axis, with perfectly conducting walls at $z=0$ and $z=l$,
therefore the transverse electric field must vanish at the boundary.

Recall that in SI units, the source-free Maxwell equations, which
our single mode field must satisfy, are
\begin{equation}
\nabla\times\mathbf{E}=\frac{\partial\mathbf{B}}{\partial t}
\end{equation}
\begin{equation}
\nabla\times\mathbf{B}=\mu_{0}\varepsilon_{0}\frac{\partial\mathbf{B}}{\partial t}
\end{equation}
\begin{equation}
\nabla\cdot\mathbf{E}=0
\end{equation}
\begin{equation}
\nabla\cdot\mathbf{B}=0
\end{equation}

We will assume that the field is polarized in the x-direction i.e.
$\mathbf{E}\left(\mathbf{r},\, t\right)=\mathbf{e}_{x}\, E_{x}\left(z,t\right)$
and hence $\mathbf{B}\left(\mathbf{r},\, t\right)=\mathbf{e}_{y}B_{y}\left(z,t\right)$.
If we identify $q\left(t\right)$ as the canonical position as defined
in the Hamiltonian formalism and similarly identify $\dot{q}\left(t\right)$
as the canonical momentum, then the solution for the components is
\begin{equation}
E_{x}\left(z,t\right)=\left(\frac{2\omega^{2}}{V\varepsilon_{0}}\right)^{1/2}q\left(t\right)\sin\left(kz\right)
\end{equation}
 and 
\begin{equation}
B_{y}\left(z,t\right)=\left(\frac{\mu_{0}\varepsilon_{0}}{k}\right)\left(\frac{2\omega^{2}}{V\varepsilon_{0}}\right)^{1/2}\dot{q}\left(t\right)\cos\left(kz\right)
\end{equation}
where the wave-number $k$ is related to the frequency $\,\omega\,$
by $\, k=\omega/c\,$. Moreover, the boundary conditions on the electric
field at the interface of the perfect conductor at $\, z=0\,$ and
$\, z=L,$ constrain the values of $k$ to be 
\begin{equation}
k=\left(\frac{n\pi}{L}\right)\;,\, n=1,2,..
\end{equation}
 and therefore the allowed frequencies are 
\begin{equation}
\omega=c\left(\frac{n\pi}{L}\right)\;,\, n=1,2,..
\end{equation}

We can invert the 2 equations giving $E_{x}$ and $B_{y}$ in terms
of the canonical position and momentum $\, q\left(t\right)\,$ and
$\,\dot{q}\left(t\right)\,$, and obtain the expressions for $\, q\left(t\right)\,$
and $\,\dot{q}\left(t\right)\,$ in terms of $\, E_{x}\,$ and $\, B_{y}\,$,
namely
\begin{equation}
q\left(t\right)=E_{x}\left(z,t\right)\left(\frac{V\varepsilon_{0}}{2\omega^{2}}\right)^{1/2}\csc\left(kz\right)
\end{equation}

\begin{equation}
\dot{q}\left(t\right)=B_{y}\left(z,t\right)\left(\frac{k}{\mu_{0}\varepsilon_{0}}\right)\left(\frac{V\varepsilon_{0}}{2\omega^{2}}\right)^{1/2}\sec\left(kz\right)
\end{equation}

From these expressions, it is apparent that the Hamiltonian for the
field is

\begin{equation}
H=\frac{1}{2}\int dV\left[\varepsilon_{0}\mathbf{E}^{2}\left(\mathbf{r},t\right)+\frac{1}{\mu_{0}}\mathbf{B}^{2}\left(\mathbf{r},t\right)\right]
\end{equation}
 Now, 
\[
\varepsilon_{0}\mathbf{E}^{2}\left(\mathbf{r},t\right)=\varepsilon_{0}\mathbf{E}\left(\mathbf{r},t\right)\cdot\mathbf{E}\left(\mathbf{r},t\right)=\varepsilon_{0}\left(\mathbf{e}_{x}\cdot\mathbf{e}_{x}\right)E_{x}^{2}\left(z,t\right)
\]
\begin{equation}
=\varepsilon_{0}E_{x}^{2}\left(z,t\right)
\end{equation}
 and similarly, 
\begin{equation}
\frac{1}{\mu_{0}}\mathbf{B}^{2}\left(\mathbf{r},t\right)=\frac{1}{\mu_{0}}B_{y}^{2}\left(z,t\right)
\end{equation}
 Therefore 
\begin{equation}
H=\frac{1}{2}\int dV\left[\varepsilon_{0}E_{x}^{2}\left(z,t\right)+\frac{1}{\mu_{0}}B_{y}^{2}\left(z,t\right)\right]
\end{equation}
 From (1) we have 
\begin{equation}
\varepsilon_{0}E_{x}^{2}\left(z,t\right)=\frac{2\omega^{2}}{V}q^{2}\left(t\right)\sin^{2}\left(kz\right)
\end{equation}
 and 
\begin{equation}
\frac{1}{\mu_{0}}B_{y}^{2}\left(z,t\right)=\frac{2}{V}p^{2}\left(t\right)\cos^{2}\left(kz\right)
\end{equation}
 Therefore (6) becomes 
\begin{equation}
H=\frac{1}{2}\int\frac{dV}{V}\left[\omega^{2}q^{2}\left(t\right)\sin^{2}\left(kz\right)+p^{2}\left(t\right)\cos^{2}\left(kz\right)\right]
\end{equation}
 Since, 
\begin{equation}
\cos^{2}x=\frac{1+\cos2x}{2}
\end{equation}
 and, 
\begin{equation}
\sin^{2}x=\frac{1-\cos2x}{2}
\end{equation}
we may write the Hamiltonian as 
\begin{equation}
H=\frac{1}{2}\int\frac{dV}{V}\left[\omega^{2}q^{2}\left(t\right)\left(1+\cos2kz\right)+p^{2}\left(t\right)\left(1-\cos2kz\right)\right]
\end{equation}
 Now the cosine terms drop out of because of the periodic boundary
conditions and therefore,
\begin{equation}
H=\frac{1}{2}\left(p^{2}+\omega^{2}q^{2}\right)
\end{equation}
and so the system is equivalent to harmonic oscillator with unit mass.
($\dot{q}\left(t\right)=p\left(t\right)$).

Now that we have the canonical momentum and canonical position, is
is relatively easy to quantize the field by replacing the variables
$H$ with $\hat{H}$ and $q\left(t\right)$ and $\dot{q}\left(t\right)$
with the hermitean (observable) operators $\hat{q}$ and $\hat{p}$,
respectively. Moreover, we must require that the observables obey
the canonical commutation relation 
\begin{equation}
\left[\hat{\, q\,},\hat{\, p\,}\right]=i\hbar\hat{I}_{n\times n}
\end{equation}
 which will write from here on out simply as44 
\begin{equation}
\left[\,\hat{q\,}\,,\hat{p}\,\right]=i\hbar
\end{equation}
 with the $n\times n$ matrix identity operator $\hat{I}_{n\times n}$
implied. Having promoted $\hat{q}$ and $\hat{p}$ to operators, we
are thereby led to the operators for the electric and magnetic fields
\begin{equation}
\hat{E_{x}}=\left(\frac{2\omega^{2}}{V\varepsilon_{0}}\right)^{1/2}\hat{q}\sin\left(kz\right)
\end{equation}
 
\begin{equation}
\hat{B_{y}}=\left(\frac{\mu_{0}\varepsilon_{0}}{k}\right)\left(\frac{2\omega^{2}}{V\varepsilon_{0}}\right)^{1/2}\hat{p}\cos\left(kz\right)
\end{equation}
 and naturally, the Hamiltonian operator becomes 
\begin{equation}
\hat{H}=\frac{1}{2}\left(\hat{p}^{2}+\omega^{2}\hat{q}^{2}\right)
\end{equation}

Now we define the non-hermitean creation, $\hat{a}^{\dagger},$ and
annihilation, $\hat{a},$ operators as follows{[}1{]}: 
\begin{equation}
\sqrt{2\hbar\omega}\hat{a}^{\dagger}=\left(\omega\hat{q}-i\hat{p}\right)
\end{equation}
\begin{equation}
\sqrt{2\hbar\omega}\hat{a}=\left(\omega\hat{q}+i\hat{p}\right)
\end{equation}
 Defining 
\begin{equation}
\mathcal{E}_{0}=\left(\hbar\omega/V\varepsilon_{0}\right)^{1/2}
\end{equation}
 and 
\begin{equation}
\mathcal{B}_{0}=\left(\mu_{0}/k\right)\left(\varepsilon_{0}\hbar\omega^{3}/V\right)^{1/2}
\end{equation}
it follows that we can write the operators for the electric and magnetic
fields as {[}1{]}:
\begin{equation}
\hat{E_{x}}\left(z,t\right)=\mathcal{E}_{0}\left(\hat{a}^{\dagger}\left(t\right)+\hat{a}\left(t\right)\right)\sin\left(kz\right)
\end{equation}
 
\begin{equation}
\hat{B_{y}}\left(z,t\right)=i\mathcal{B}_{0}\left(\hat{a}^{\dagger}\left(t\right)-\hat{a}\left(t\right)\right)\cos\left(kz\right)
\end{equation}

From now on we will suppress hats, $\,\hat{}\;,$ on operators and
just write $\hat{a}^{\dagger}=a^{\dagger},$ $\hat{a}=a,$ $\hat{E_{x}}=E_{x},$
etc. The benefit of working with creation and annihilation operators
is that we are allowed to utilize the simplicity of their algebra.
\begin{equation}
\left[\: a\,,\, a^{\dagger}\,\right]=aa^{\dagger}-a^{\dagger}a=1
\end{equation}
 These commutation relations allow us to write the Hamiltonian operator
as {[}2{]} 
\begin{equation}
H=\hbar\omega\left(a^{\dagger}a+\frac{1}{2}\right)
\end{equation}
 In the Heisenberg representation, a general operator $\hat{O}$ will
obey Heisenberg's equation of motion {[}2{]} 
\begin{equation}
\frac{d\hat{O}}{dt}=\frac{\partial\hat{O}}{dt}+\frac{i}{\hbar}\left[\, H\,,\,\hat{O}\,\right]
\end{equation}
Which in the case that $\hat{O}$ does not depend explicitly on the
time coordinate, becomes 
\begin{equation}
\frac{d\hat{O}}{dt}=\frac{i}{\hbar}\left[\, H\,,\,\hat{O}\,\right]
\end{equation}
 Therefore for the creation and annihilation operators we have the
following time evolution equations{[}1{]}
\begin{equation}
\frac{da^{\dagger}}{dt}=i\omega a^{\dagger}
\end{equation}
 
\begin{equation}
\frac{da}{dt}=-i\omega a
\end{equation}
 which implies that
\begin{equation}
a^{\dagger}\left(t\right)=a^{\dagger}\left(0\right)e^{i\omega t}
\end{equation}
 and 
\begin{equation}
a\left(t\right)=a\left(0\right)e^{-i\omega t}
\end{equation}

We may expand $e^{-i\omega t}$ as 
\begin{equation}
e^{-i\omega t}=1-i\omega t-\frac{\omega^{2}t^{2}}{2!}+i\frac{\omega^{3}t^{3}}{3!}+...
\end{equation}
 which allows us to write $a\left(t\right)$ as 
\begin{equation}
a\left(t\right)=a\left(0\right)\left(1-i\omega t-\frac{\omega^{2}t^{2}}{2!}+i\frac{\omega^{3}t^{3}}{3!}+...\right)
\end{equation}
 A useful combination of operators will be $a^{\dagger}a=n$ , a combination
known as the number operator. If applied to the $n^{th}$ eigenstate
of the Hamiltonian $\left|n\right\rangle $ (we will later come to
identify $\left|n\right\rangle $ as the $n$ photon state), 
\begin{equation}
a^{\dagger}a\left|n\right\rangle =n\left|n\right\rangle 
\end{equation}
this operator gives the value $n$ of the eigenstate occupied. The
energy eigenvalue problem can then be written as 
\begin{equation}
H\left|n\right\rangle =\hbar\omega\left(a^{\dagger}a+\frac{1}{2}\right)\left|n\right\rangle 
\end{equation}
\begin{equation}
=\hbar\omega\left(n+\frac{1}{2}\right)\left|n\right\rangle =E_{n}\left|n\right\rangle 
\end{equation}
 Therefore 
\begin{equation}
E_{n}=\hbar\omega\left(n+\frac{1}{2}\right).
\end{equation}
Where $E_{0}$ is the ground state energy, since $0$ is the lowest
value which may be taken by $n$ as can be seen from acting on the
state $\left|0\right\rangle $ with the annihilation operator $a$.
\begin{equation}
a\left|0\right\rangle =0
\end{equation}
 Since 
\begin{equation}
a\left|n\right\rangle =\sqrt{n}\left|n-1\right\rangle .
\end{equation}
and 
\begin{equation}
a^{\dagger}\left|n\right\rangle =\sqrt{n+1}\left|n+1\right\rangle .
\end{equation}
 It follows that any arbitrary eigenstate $\left|n\right\rangle $
can be written in terms of the vacuum state as (e.g. {[}1{]})
\begin{equation}
\left|n\right\rangle =\frac{\left(a^{\dagger}\right)^{n}}{\sqrt{n!}}\left|0\right\rangle .
\end{equation}
 The states $\left|n\right\rangle $ form a complete basis for the
Hamiltonian $H$, and are orthonormal 
\begin{equation}
\left\langle m\right|\left|n\right\rangle =\delta_{mn}
\end{equation}
 The non vanishing matrix elements of the creation and annihilation
operators are 
\begin{equation}
\left\langle n-1\right|a\left|n\right\rangle =\sqrt{n}
\end{equation}
 
\begin{equation}
\left\langle n+1\right|a^{\dagger}\left|n\right\rangle =\sqrt{n+1}
\end{equation}
. 
\end{doublespace}

\begin{doublespace}

\subsection{Quantum fluctuations of the single mode field.}
\end{doublespace}

\begin{doublespace}
Recall the operator for the electric field given in (8), 
\begin{equation}
\hat{E_{x}}\left(z,t\right)=\mathcal{E}_{0}\left(\hat{a}^{\dagger}\left(t\right)+\hat{a}\left(t\right)\right)\sin\left(kz\right).
\end{equation}
The eigenstates of the Hamiltonian $\left|n\right\rangle $, do not
form a basis for the operator $\hat{E_{x}}\left(z,t\right)$. This
is implied by the fact that the number operator $n=a^{\dagger}a$,
which does commute with the Hamiltonian, does not commute with the
electric field operator $E_{x}$. 
\end{doublespace}

To see this, let us first calculate the average field $\left\langle E_{x}\right\rangle $
\begin{equation}
\left\langle E_{x}\right\rangle =\left\langle n\right|\hat{E_{x}}\left(z,t\right)\left|n\right\rangle =\mathcal{E}_{0}\left[\left\langle n\right|a\left|n\right\rangle +\left\langle n\right|a^{\dagger}\left|n\right\rangle \right]\sin\left(kz\right)
\end{equation}
 which, by (10) and (11), become 
\begin{equation}
=\mathcal{E}_{0}\left[0+0\right]\sin\left(kz\right)=\left\langle \hat{E_{x}}\right\rangle =0
\end{equation}
that is, the average field is zero.

\begin{doublespace}
The energy density of the field $\mathbf{E}=\mathbf{e}_{x}E_{x}$
is proportional to the mean square of $E_{x}$, {[}1{]} 
\begin{equation}
\left\langle E_{x}^{2}\right\rangle =2\mathcal{E}_{0}^{2}\sin^{2}\left(kz\right)\left(n+\frac{1}{2}\right)
\end{equation}
 The \emph{variance }is defined as {[}1{]} 
\begin{equation}
\left\langle (\Delta E_{x})^{2}\right\rangle =\left\langle E_{x}^{2}\right\rangle -\left\langle \hat{E_{x}}\right\rangle ^{2}
\end{equation}
 i.e, its the mean square of the standard deviation, which for the
eigenstate $\left|\varphi_{n}\right\rangle $ becomes 
\begin{equation}
\Delta E_{x}=\sqrt{\left\langle E_{x}^{2}\right\rangle -\left\langle \hat{E_{x}}\right\rangle ^{2}}=\sqrt{2\mathcal{E}_{0}^{2}\sin^{2}\left(kz\right)\left(n+\frac{1}{2}\right)}
\end{equation}
that is, 
\begin{equation}
\Delta E_{x}=\sqrt{2}\mathcal{E}_{0}\sin\left(kz\right)\left(n+\frac{1}{2}\right)^{1/2}
\end{equation}
 It is interesting to note that even for $n=0$ we have 
\begin{equation}
\Delta E_{x}=\sqrt{2}\mathcal{E}_{0}\sin\left(kz\right)\frac{1}{\sqrt{2}}=\mathcal{E}_{0}\sin\left(kz\right)
\end{equation}
 these are called the \emph{vacuum fluctuations }of the field, since
they correspond to the eigenstate of the vacuum $\left|0\right\rangle $,
the state with zero photons. {[}1{]} 

In the case of the electromagnetic field confined to a 1D cavity,
the eigenstates of the Hamiltonian, namely$\left|n\right\rangle $,
correspond to states of photon number $n$ . One important fact is
that the number operator $n=a^{\dagger}a$ and the electric field
operator $E_{x}$ do not commute, 
\begin{equation}
\left[\, n\,,\, E_{x}\,\right]=\mathcal{E}_{0}\sin\left(kz\right)\left(a^{\dagger}-a\right)
\end{equation}
The generalized uncertainty relations state that for any two operators
$A$ and $B$ satisfying $\left[\, A\,,\, B\,\right]=C$ , it follows
that the product of the uncertainties of $A$ with that of $B$ obey
the inequality 
\begin{equation}
\Delta A\:\triangle B\geq\frac{1}{2}\left|\left\langle C\right\rangle \right|
\end{equation}
It follows therefore, that the number operator and the electric field
obey the following uncertainty relations 
\begin{equation}
\Delta n\Delta E_{x}\geq\frac{1}{2}\mathcal{E}_{0}\left|\sin\left(kz\right)\right|\left|\left\langle a^{\dagger}-a\right\rangle \right|
\end{equation}
 This implies a number-phase uncertainty relation {[}1{]} 
\begin{equation}
\Delta n\:\triangle\phi\geq1
\end{equation}
 where $0<\phi<2\pi$ is the \emph{phase angle }associated with the
creation and annihilation operators. In fact, it will turn out, that
the situation is not actually quite that simple. It turns out to be
a very slippery task to define a unique phase operator, and in fact
is not possible in general {[}5{]} ,{[}6{]}, {[}7{]}. It can be shown,
however, that for proper definitions of the phase,namely those given
in {[}4{]}, that the photon number states $\left|\varphi_{n}\right\rangle $
have a uniform phase distribution $\left(\Delta\phi/\Delta n\right)\approx constant$
for $0\leq\phi\leq2\pi$. For more on the number phase uncertainty
relations see {[}3{]}, {[}4{]} and the references given in {[}1{]}
.
\end{doublespace}

\begin{doublespace}

\subsection{Multimode fields.}
\end{doublespace}

\begin{doublespace}
In free space in the absence of any sources, the source free Maxwell
equations are still valid, (joking but obviously true). We write the
electric and magnetic fields in terms of the vector potential $\mathbf{A}\left(\mathbf{r},t\right)$
which satisfies the wave equation {[}1{]} 
\begin{equation}
\nabla^{2}\mathbf{A}-\frac{1}{c^{2}}\frac{\partial^{2}\mathbf{A}}{\partial t^{2}}=0
\end{equation}
 and we choose the Coulomb gauge condition(which will become useful
later on) 
\begin{equation}
\nabla\cdot\mathbf{A}=0
\end{equation}
 The electric field is then given by{[}1{]} 
\begin{equation}
\mathbf{E}=-\frac{\partial\mathbf{A}}{\partial t}
\end{equation}
and the magnetic field is 
\begin{equation}
\mathbf{B}=\nabla\times\mathbf{A}
\end{equation}

As long as $L\gg\frac{1}{k}$ , we can model free space as cubic cavity,
with sides of length $L$, therefore we may impose periodic boundary
conditions on the faces of the cube{[}1{]}. This allows us to deal
with the mathematically simpler case of having a denumerably infinite
set of normal modes, rather than a non-denumerably infinite set of
modes {[}1{]}. We require plane waves in the $x_{i}$ direction, where
$\left(i=1,2,3\right)$ and $\left(x_{1}=x,\, x_{2}=y,\, x_{3}=z\right)$,
to satisfy the condition 
\begin{equation}
e^{ik_{x_{i}}x_{i}}=e^{ik_{x_{i}}\left(x_{i}+L\right)}
\end{equation}
which leads the following conditions for the direction numbers $k_{x_{i}}$
\begin{equation}
k_{x}=\left(\frac{2\pi}{L}\right)m_{x}
\end{equation}
\begin{equation}
k_{y}=\left(\frac{2\pi}{L}\right)m_{y}
\end{equation}
\begin{equation}
k_{z}=\left(\frac{2\pi}{L}\right)m_{z}
\end{equation}
where 
\begin{equation}
m_{x}=m_{y}=m_{z}=0,\pm1,\pm2,...
\end{equation}
 Now, the wave vector 
\begin{equation}
\mathbf{k}=\left(k_{x},k_{y},k_{z}\right)=\frac{2\pi}{L}\left(m_{x},m_{y},m_{z}\right)
\end{equation}
 Moreover, $k=\left\Vert \mathbf{k}\right\Vert =\sqrt{\mathbf{k}\cdot\mathbf{k}}=\omega_{k}/c$
. Distinct normal modes of the fields are specified by distinct sets
of integers $\left(m_{x},m_{y},m_{z}\right)$. Therefore, the total
numbers of modes in the interval $\left(\Delta m_{x},\Delta m_{y},\Delta m_{z}\right)$
is {[}1{]} 
\begin{equation}
\Delta m=\Delta m_{x}\Delta m_{y}\Delta m_{z}=2\left(\frac{L}{2\pi}\right)^{3}\Delta k_{x}\Delta k_{y}\Delta k_{z}
\end{equation}
 taking into account a factor of 2 for the two independent polarizations.
In the limit that $L\rightarrow\infty$, $\triangle m\rightarrow dm$
and we have $\left(V=L^{3}\right)$ 
\begin{equation}
dm=dm_{x}dm_{y}dm_{z}=\left(\frac{V}{4\pi^{3}}\right)dk_{x}dk_{y}dk_{z}
\end{equation}
 going to spherical coordinates this is 
\begin{equation}
\mathbf{k}=\left(k_{x},k_{y},k_{z}\right)=k\left(\sin\theta\cos\phi,\sin\theta\sin\phi,\cos\theta\right)
\end{equation}
and therefore 
\begin{equation}
dm=\left(\frac{V}{4\pi^{3}}\right)k^{2}dk\, d\Omega
\end{equation}
 or using $k=\omega_{k}/c$ we can write this as 
\begin{equation}
dm=\left(\frac{V}{4\pi^{3}}\right)\frac{\omega_{k}^{2}k}{c^{3}}d\omega_{k}\, d\Omega
\end{equation}
Integrating over the solid angle $\Omega$, we obtain 
\begin{equation}
dm=\frac{V}{\pi^{2}}k^{2}dk=V\rho_{k}dk
\end{equation}
Where $\rho_{k}=k^{2}/\pi^{2}$. We may also write this for $d\omega_{k}$
as 
\begin{equation}
dm=V\frac{\omega_{k}^{2}}{\pi^{2}c^{3}}d\omega_{k}=V\rho\left(\omega_{k}\right)d\omega_{k}
\end{equation}
 for which 
\begin{equation}
\rho\left(\omega_{k}\right)=\omega_{k}^{2}/\left(\pi^{2}c^{3}\right)
\end{equation}
. 

Having ``pasted'' the cubic grid on our space, we may proceed to
expand the vector potential as 
\begin{equation}
\mathbf{A}\left(\mathbf{r},t\right)=\sum_{\mathbf{k},s}\mathbf{e}_{\mathbf{k}s}\left[A_{\mathbf{k}s}\left(t\right)e^{i\mathbf{k\cdot r}}+A_{\mathbf{k}s}^{*}\left(t\right)e^{-i\mathbf{k\cdot r}}\right]
\end{equation}
 where $A_{\mathbf{k}s}\in\mathbb{C}$ is the amplitude of the field
and $\mathbf{e}_{\mathbf{k}s}\in\mathbb{R}$ is a polarization vector
{[}1{]}. Moreover the sum over $\mathbf{k}$ is the sum over the distinct
sets of integers $\left(m_{x},m_{y},m_{z}\right)$, and the sum over
$s$ is the sum over the two polarization directions {[}1{]}, which
must obey the orthonormality relations 
\begin{equation}
\mathbf{e}_{\mathbf{k}s}\cdot\mathbf{e}_{\mathbf{k}s'}=\delta_{\mathbf{kk'}}\delta_{ss'}
\end{equation}
 The Coulomb gauge condition requires that $\mathbf{k}\cdot\mathbf{e}_{\mathbf{k}s}=0$,
which is known as the \emph{transversality }condition {[}1{]}.

The wave equation and the Coulomb gauge lead to the following relations
for the amplitudes $A_{\mathbf{k}s}$: 
\begin{equation}
\frac{d^{2}A_{\mathbf{k}s}}{dt^{2}}+\omega_{k}^{2}A_{\mathbf{k}s}=0
\end{equation}
 The solution to this differential equation is 
\begin{equation}
A_{\mathbf{k}s}\left(t\right)=A_{\mathbf{k}s}e^{-i\omega_{k}t}
\end{equation}
 $\left(A_{\mathbf{k}s}\left(0\right)\equiv A_{\mathbf{k}s}\right)$.
Thus, the electric and magnetic fields become {[}1{]} 
\begin{equation}
\mathbf{E}\left(\mathbf{r},t\right)=i\sum_{\mathbf{k},s}\omega_{k}\mathbf{e}_{\mathbf{k}s}\left[A_{\mathbf{k}s}e^{i\left(\mathbf{k\cdot r}-\omega_{k}t\right)}+A_{\mathbf{k}s}^{*}\left(t\right)e^{-i\left(\mathbf{k\cdot r}-\omega_{k}t\right)}\right]
\end{equation}
 
\begin{equation}
\mathbf{B}\left(\mathbf{r},t\right)=\frac{i}{c}\sum_{\mathbf{k},s}\omega_{k}\left(\frac{\mathbf{k}}{\left|\mathbf{k}\right|}\times\mathbf{e}_{\mathbf{k}s}\right)\left[A_{\mathbf{k}s}e^{i\left(\mathbf{k\cdot r}-\omega_{k}t\right)}+A_{\mathbf{k}s}^{*}\left(t\right)e^{-i\left(\mathbf{k\cdot r}-\omega_{k}t\right)}\right]
\end{equation}

The energy of the field is 
\begin{equation}
H=\frac{1}{2}\int dV\left[\varepsilon_{0}\mathbf{E}\cdot\mathbf{E}+\frac{1}{\mu_{0}}\mathbf{B}\cdot\mathbf{B}\right]
\end{equation}
 Now {[}1{]}, 
\begin{equation}
\left(\frac{\mathbf{k}}{\left|\mathbf{k}\right|}\times\mathbf{e}_{\mathbf{k}s}\right)\cdot\left(\frac{\mathbf{k}}{\left|\mathbf{k}\right|}\times\mathbf{e}_{\mathbf{k}s'}\right)=\delta_{ss'}
\end{equation}
 and
\begin{equation}
\left(\frac{\mathbf{k}}{\left|\mathbf{k}\right|}\times\mathbf{e}_{\mathbf{k}s}\right)\cdot\left(\frac{\mathbf{k}}{\left|\mathbf{k}\right|}\times\mathbf{e}_{\mathbf{k}s'}\right)=-\mathbf{e}_{\mathbf{k}s}\cdot\mathbf{e}_{\mathbf{-k}s'}
\end{equation}

Taking our periodic boundary conditions into account, we have 
\begin{equation}
\int e^{\pm i\left(\mathbf{k-k'}\right)\cdot\mathbf{r}}dV=\delta_{kk'}V
\end{equation}
 Therefore, the contribution to $H$ from the electric field is 
\begin{equation}
\frac{1}{2}\int dV\left[\varepsilon_{0}\mathbf{E}\cdot\mathbf{E}\right]=\varepsilon_{0}V\sum_{\mathbf{k},s}\omega_{k}^{2}A_{\mathbf{k}s}A_{\mathbf{k}s}^{*}-R
\end{equation}
 The contribution from the magnetic field is 
\begin{equation}
\frac{1}{2}\int dV\frac{1}{\mu_{0}}\mathbf{B}\cdot\mathbf{B}=\varepsilon_{0}V\sum_{\mathbf{k},s}\omega_{k}^{2}A_{\mathbf{k}s}A_{\mathbf{k}s}^{*}+R
\end{equation}
 where, 
\begin{equation}
R=\frac{1}{2}\varepsilon_{0}V\sum_{\mathbf{k},s}\omega_{k}^{2}\mathbf{e}_{\mathbf{k}s}\cdot\mathbf{e}_{-\mathbf{k}s'}\left[A_{\mathbf{k}s}\left(t\right)A_{-\mathbf{k}s'}\left(t\right)+A_{\mathbf{k}s}^{*}\left(t\right)A_{-\mathbf{k}s'}^{*}\left(t\right)\right]
\end{equation}
 Therefore, the total energy in the field is 
\begin{equation}
H=2\varepsilon_{0}V\sum_{\mathbf{k},s}\omega_{k}^{2}A_{\mathbf{k}s}\left(t\right)A_{\mathbf{k}s}^{*}\left(t\right)
\end{equation}
 but since $A_{\mathbf{k}s}\left(t\right)=A_{\mathbf{k}s}e^{-i\omega_{k}t}$,
this may be written as 
\begin{equation}
H=2\varepsilon_{0}V\sum_{\mathbf{k},s}\omega_{k}^{2}A_{\mathbf{k}s}A_{\mathbf{k}s}^{*}
\end{equation}

\end{doublespace}

\begin{doublespace}

\subsection{Quantization of the multimode field.}
\end{doublespace}

\begin{doublespace}
We may quantize the field by introducing the canonical position and
momentum operators $q_{\mathbf{k}s}$ and $p_{\mathbf{k}s}$, respectively
through the definitions 
\begin{equation}
A_{\mathbf{k}s}=\frac{1}{2\omega_{k}\sqrt{\varepsilon_{0}V}}\left[\omega_{k}q_{\mathbf{k}s}+ip_{\mathbf{k}s}\right]
\end{equation}
\begin{equation}
A_{\mathbf{k}s}^{*}=\frac{1}{2\omega_{k}\sqrt{\varepsilon_{0}V}}\left[\omega_{k}q_{\mathbf{k}s}-ip_{\mathbf{k}s}\right]
\end{equation}
 in which case the Hamiltonian becomes 
\begin{equation}
H=\frac{1}{2}\sum_{\mathbf{k},s}\left(p_{\mathbf{k}s}^{2}+\omega_{k}^{2}q_{\mathbf{k}s}^{2}\right)
\end{equation}
as it should. 

The canonical variables obey the canonical commutation relations 
\begin{equation}
\left[\, q_{\mathbf{k}s}\,,\, q_{\mathbf{k}'s'}\right]=0=\left[p_{\mathbf{k}s},p_{\mathbf{k}'s'}\right]
\end{equation}
\begin{equation}
\left[q_{\mathbf{k}s}\,,\, p_{\mathbf{k}'s'}\right]=i\hbar\delta_{\mathbf{k}\mathbf{k'}}\delta_{ss'}
\end{equation}
 Just as we did for the single mode field, we may define the creation
and annihilation operators for the multimode fields 
\begin{equation}
\sqrt{2\hbar\omega_{k}}a_{\mathbf{k}s}=\omega_{k}q_{\mathbf{k}s}+ip_{\mathbf{k}s}
\end{equation}
\begin{equation}
\sqrt{2\hbar\omega_{k}}a_{\mathbf{k}s}^{\dagger}=\omega_{k}q_{\mathbf{k}s}-ip_{\mathbf{k}s}
\end{equation}

The creation and annihilation operators obey the following commutation
relations 
\begin{equation}
\left[a_{\mathbf{k}s}\,,\, a_{\mathbf{k'}s'}\right]=\left[a_{\mathbf{k}s}^{\dagger}\,,\, a_{\mathbf{k}'s'}^{\dagger}\right]=0
\end{equation}
 
\begin{equation}
\left[a_{\mathbf{k}s}\,,\, a_{\mathbf{k'}s'}^{\dagger}\right]=\delta_{\mathbf{k}\mathbf{k'}}\delta_{ss'}\delta\left(\mathbf{k}^{'}-\mathbf{k}\right)
\end{equation}
 Just as was the case for the single mode field, the number operator
for the mode $\mathbf{k}s$ is $n_{\mathbf{k}s}=a_{\mathbf{k}s}^{\dagger}a_{\mathbf{k}s}$,
and the Hamiltonian is 
\begin{equation}
H=\sum_{\mathbf{k},s}\hbar\omega_{k}\left(a_{\mathbf{k}s}^{\dagger}a_{\mathbf{k}s}+\frac{1}{2}\right)
\end{equation}
\begin{equation}
=\sum_{\mathbf{k},s}\hbar\omega_{k}\left(n_{\mathbf{k}s}+\frac{1}{2}\right)
\end{equation}

Each mode is independent of all the rest and has the eigenstates $\left|n_{\mathbf{k}s}\right\rangle $.
If we let $j$ denote the $j^{th}$ mode $\mathbf{k}_{j}s_{j}$, then
we may write the $n^{th}$ photon number state of the $j^{th}$ mode
as 
\begin{equation}
\left|\left\{ n_{j}\right\} \right\rangle =\prod_{j}\frac{\left(a_{j}^{\dagger}\right)^{n_{j}}}{\sqrt{n_{j}!}}\left|\varphi_{0}^{\left(j\right)}\right\rangle 
\end{equation}
 The energy eigenvalue equation is then 
\begin{equation}
H\left|\left\{ n_{j}\right\} \right\rangle =E\left|\left\{ n_{j}\right\} \right\rangle 
\end{equation}
 where {[}1{]} 
\begin{equation}
E=\sum_{j}\hbar\omega_{j}\left(n_{j}+\frac{1}{2}\right)
\end{equation}
 A multimode photon state is the tensor product of all of the individual
mode number states, that is 
\begin{equation}
\left|n_{1},n_{2},n_{3}...\right\rangle =\left|n_{1}\right\rangle \otimes\left|n_{2}\right\rangle \otimes\left|n_{3}\right\rangle \otimes...
\end{equation}
\begin{equation}
=\left|\left\{ n_{j}\right\} \right\rangle 
\end{equation}
The number states are orthogonal, that is 
\begin{equation}
\left\langle \left\{ n_{j'}\right\} \right|\left|\left\{ m_{j}\right\} \right\rangle =\prod_{j,j'}\delta_{n_{j}m_{j}}\delta_{jj'}
\end{equation}
 The multimode vacuum state is 
\begin{equation}
\left|\left\{ 0_{j}\right\} \right\rangle =\left|0\right\rangle \otimes\left|0\right\rangle \otimes\left|0\right\rangle \otimes...
\end{equation}
The action of the creation and annihilation operators on $j^{th}$
mode of the multimode photon number state are given by 
\begin{equation}
a_{j}\left|\left\{ n_{j}\right\} \right\rangle =\sqrt{n_{j}}\left|\left\{ \left(n-1\right)_{j}\right\} \right\rangle 
\end{equation}
 
\begin{equation}
a_{j}^{\dagger}\left|\left\{ n_{j}\right\} \right\rangle =\sqrt{n_{j}+1}\left|\left\{ \left(n+1\right)_{j}\right\} \right\rangle 
\end{equation}
 Quantization requires that the amplitudes $A_{\mathbf{k}s}$ become
the operators: 
\begin{equation}
\hat{A}_{\mathbf{k}s}=\left(\frac{\hbar}{2\omega_{k}\varepsilon_{0}V}\right)^{1/2}a_{\mathbf{k}s}
\end{equation}
 Which therefore allows us to define a vector potential operator as
well as electric and magnetic field operators, which are, respectively,

\begin{equation}
\mathbf{\hat{A}}\left(\mathbf{r},t\right)=\sum_{\mathbf{k},s}\left(\frac{\hbar}{2\omega_{k}\varepsilon_{0}V}\right)^{1/2}\mathbf{e}_{\mathbf{k}s}\left[a_{\mathbf{k}s}e^{i\left(\mathbf{k\cdot r}-\omega_{k}t\right)}+a_{\mathbf{k}s}^{\dagger}e^{-i\left(\mathbf{k\cdot r}-\omega_{k}t\right)}\right]
\end{equation}
\begin{equation}
\mathbf{\hat{E}}\left(\mathbf{r},t\right)=i\sum_{\mathbf{k},s}\left(\frac{\hbar\omega_{k}}{2\varepsilon_{0}V}\right)^{1/2}\mathbf{e}_{\mathbf{k}s}\left[a_{\mathbf{k}s}e^{i\left(\mathbf{k\cdot r}-\omega_{k}t\right)}-a_{\mathbf{k}s}^{\dagger}e^{-i\left(\mathbf{k\cdot r}-\omega_{k}t\right)}\right]
\end{equation}
 
\begin{equation}
\mathbf{\hat{B}}\left(\mathbf{r},t\right)=\frac{i}{c}\sum_{\mathbf{k},s}\omega_{k}\left(\frac{\mathbf{k}}{\left|\mathbf{k}\right|}\times\mathbf{e}_{\mathbf{k}s}\right)\left(\frac{\hbar\omega_{k}}{2\varepsilon_{0}V}\right)^{1/2}\left[a_{\mathbf{k}s}e^{i\left(\mathbf{k\cdot r}-\omega_{k}t\right)}-a_{\mathbf{k}s}^{\dagger}e^{-i\left(\mathbf{k\cdot r}-\omega_{k}t\right)}\right]
\end{equation}
 where the operators $a_{\mathbf{k}s}=a_{\mathbf{k}s}\left(0\right)$
form a basis of the \emph{Heisenberg representation}. 

The time dependent creation and annihilation operators are, respectively
\begin{equation}
a_{\mathbf{k}s}^{\dagger}\left(t\right)=a_{\mathbf{k}s}^{\dagger}\left(0\right)e^{i\omega_{k}t}
\end{equation}
\begin{equation}
a_{\mathbf{k}s}\left(t\right)=a_{\mathbf{k}s}\left(0\right)e^{-i\omega_{k}t}
\end{equation}
 It can be seen that the magnetic field is weaker than the electric
field by a factor of $1/c$, which is what we should expect and is
a reassuring sign we haven\textquoteright{}t gone off track. The magnetic
field couples to the spin magnetic moment of the electrons which is
negligible for the aspects of quantum optics which we will investigate.

Some interesting things to note are:
\end{doublespace}
\begin{itemize}
\begin{doublespace}
\item A single mode plane wave has electric field components given by 
\begin{equation}
\mathbf{\hat{E}}\left(\mathbf{r},t\right)=i\left(\frac{\hbar\omega}{2\varepsilon_{0}V}\right)^{1/2}\mathbf{e}_{x}\left[ae^{i\left(\mathbf{k\cdot r}-\omega t\right)}-a^{\dagger}e^{-i\left(\mathbf{k\cdot r}-\omega t\right)}\right]
\end{equation}
 
\item Quantum optics often works in the domain of optical radiation, whose
wavelength $\lambda$ is on the order $10^{3}\textrm{\AA}$.Therefore
in these situations we may approximate 
\begin{equation}
e^{\pm i\mathbf{k\cdot r}}\approx1\pm i\mathbf{k\cdot r}
\end{equation}
 since, in these situations it is true that 
\begin{equation}
\frac{\lambda}{2\pi}=\frac{1}{\left|\mathbf{k}\right|}\gg\left|\mathbf{r}_{atom}\right|
\end{equation}
 and hence the electric field can be expanded as 
\begin{equation}
\mathbf{\hat{E}}\left(\mathbf{r},t\right)\approx\mathbf{\hat{E}}\left(t\right)=i\left(\frac{\hbar\omega}{2\varepsilon_{0}V}\right)^{1/2}\mathbf{e}_{x}\left[ae^{-i\omega t}-a^{\dagger}e^{i\omega t}\right]
\end{equation}
 which is known as the \emph{dipole approximation }.\emph{ }\end{doublespace}

\end{itemize}
\begin{doublespace}

\section{Thermal Modes.}
\end{doublespace}

\begin{doublespace}
Consider a single mode field in thermodynamic equilibrium with the
walls of a cavity of absolute temperature $T$. The density operator
$\rho$ for the system is {[}2{]} 
\begin{equation}
\rho=Z^{-1}e^{-H/kT}
\end{equation}
 where $H$ is the Hamiltonian, $k$ is Boltzmann's constant and $Z$
is called the \emph{partition function, }and it is introduced as a
normalization factor, in order that trace of $\rho$ be one{[}2{]}.
That is 
\begin{equation}
Z=\mathrm{Tr}\left\{ e^{-H/kT}\right\} .
\end{equation}
 The density matrix for the system is diagonal in the Hamiltonian
eigenbasis $\left|\varphi_{n}\right\rangle $ since it is in thermodynamic
equilibrium. The diagonal matrix components, gives the \emph{population
}of the stationary state $\left|\varphi_{n}\right\rangle $, which
in this case are all the same. They are: 
\begin{equation}
\rho_{th}=\rho_{nn}=\left\langle n\right|Z^{-1}e^{-H/kT}\left|n\right\rangle 
\end{equation}
\begin{equation}
=Z^{-1}e^{-E_{n}/kT}.
\end{equation}
 Since $E_{n}=\hbar\omega\left(n+\frac{1}{2}\right)$, the partition
function becomes{[}1{]} 
\begin{equation}
Z=\exp\left(-\hbar\omega/2kT\right)\sum_{n}\exp\left(-\hbar\omega n/2kT\right)
\end{equation}
 Since , $\exp\left(-\hbar\omega/kT\right)<1$ we may sum the series
\begin{equation}
\sum_{n}\exp\left(-\hbar\omega/k_{B}T\right)=\frac{1}{1-\exp\left(--\hbar\omega/kT\right)}
\end{equation}
 Therefore 
\begin{equation}
Z=\frac{\exp\left(-\hbar\omega/kT\right)}{1-\exp\left(--\hbar\omega/kT\right)}
\end{equation}

The off-diagonal terms vanish, hence there are no \emph{coherences}
between stationary states, and 
\begin{equation}
\rho_{nm}=\left\langle n\right|Z^{-1}e^{-H/kT}\left|m\right\rangle 
\end{equation}
\begin{equation}
=Z^{-1}e^{-E_{m}/kT}\left\langle n\right|\left|m\right\rangle =0.
\end{equation}

We observe that in thermodynamic equilibrium, the populations of the
stationary states decrease exponentially with the energy. Since there
are no coherences, the system in this case may be considered to be
a statistical mixture of the states $\left|n\right\rangle $. 

The probability that the thermal mode is in the $n^{th}$ thermally
excited state is 
\begin{equation}
P_{n}=\left\langle n\right|\rho_{th}\left|n\right\rangle 
\end{equation}
\begin{equation}
=\frac{\exp\left(-E_{n}/kT\right)}{\sum_{n}\exp\left(-E_{n}/kT\right)}
\end{equation}
 The density operator may be written as{[}1{]} 
\begin{equation}
\rho_{th}=\sum_{n'=0}^{\infty}\sum_{n=0}^{\infty}\left|n'\right\rangle \left\langle n\right|\rho_{th}\left|n\right\rangle \left\langle n\right|
\end{equation}
 
\begin{equation}
=\frac{1}{Z}\sum_{n=0}^{\infty}\exp\left(-E_{n}/kT\right)\left|n\right\rangle \left\langle n\right|
\end{equation}
\begin{equation}
=\sum_{n=0}^{\infty}P_{n}\left|n\right\rangle \left\langle n\right|
\end{equation}
 The average photon number of the thermal field is {[}1{]}
\begin{equation}
n=\mathrm{Tr}\left(n\rho_{th}\right)=\frac{1}{\exp\left(\hbar\omega/kT\right)-1}
\end{equation}
 from which it follows that for $kT\gg\hbar\omega\rightarrow$ 
\begin{equation}
n\approx kT/\hbar\omega
\end{equation}
While for $\hbar\omega\gg kT\rightarrow$ 
\begin{equation}
n\approx\hbar\omega/kT
\end{equation}

\end{doublespace}

\begin{doublespace}

\section{The interaction of atoms and electromagnetic waves }
\end{doublespace}

\begin{doublespace}
The Hamiltonian for a system consisting an electron bound to an atom
in the presence of external fields is 
\begin{equation}
H\left(\mathrm{r},t\right)=\frac{1}{2m}\left[\mathbf{P}+e\mathbf{A}\left(\mathrm{r},t\right)\right]^{2}-e\Phi\left(\mathrm{r},t\right)+V\left(r\right)
\end{equation}
 The gauge invariant electric and magnetic fields are given by 
\begin{equation}
\mathbf{E}\left(\mathbf{r},t\right)=-\nabla\Phi\left(\mathrm{r},t\right)-\frac{\partial\mathbf{A}}{\partial t}
\end{equation}
\begin{equation}
\mathbf{B}\left(\mathbf{r},t\right)=\nabla\times\mathbf{A}\left(\mathrm{r},t\right)
\end{equation}
 Gauge invariance means that these fields are invariant under the
gauge transformations
\begin{equation}
\Phi^{'}\left(\mathrm{r},t\right)=\Phi\left(\mathrm{r},t\right)-\frac{\partial\chi\left(\mathrm{r},t\right)}{\partial t}
\end{equation}
\begin{equation}
\mathbf{A}^{'}\left(\mathrm{r},t\right)=\mathbf{A}\left(\mathrm{r},t\right)+\nabla\chi\left(\mathrm{r},t\right)
\end{equation}
 Therefore the equation governing the time evolution of the system
in the Schrodinger representation, is the Schrodinger equation 
\begin{equation}
H\left(\mathrm{r},t\right)\Psi\left(\mathrm{r},t\right)=i\hbar\frac{\partial\Psi\left(\mathrm{r},t\right)}{\partial t}.
\end{equation}
In quantum mechanics all operators are invariant under obey a global
$U\left(1\right)$ similarity transformation; for some operator $A$,$A^{'}=UAU^{\dagger}$,
where $U$ is some unitary operator. Moreover, all state vectors are
invariant under multiplication by a common $U$. This essentially
means that given some quantum mechanical representation of system
with all of its operators, and states, etc., we may obtain an equivalent
description of that same system if we simultaneously transform all
of the states and operators of the theory in the manner prescribed
above. The resulting transformed theory will lead to all of the same
results as the original theory. The usefulness of this fact is that
a particular operator may take on a more tractable form in the transformed
theory. Therefore, we may exploit this fact to simplify the Hamiltonian.
It will prove useful for us to define the unitary operator $R$ which
takes us to another representation $\Psi^{'}\left(\mathrm{r},t\right)$
of the eigenstate $\Psi\left(\mathrm{r},t\right)$, by the action
of $R$ on $\Psi\left(\mathrm{r},t\right)$, namely 
\begin{equation}
R\Psi\left(\mathrm{r},t\right)=\Psi^{'}\left(\mathrm{r},t\right).
\end{equation}
The transformed Hamiltonian obeys its own Schrodinger equation 
\begin{equation}
H^{'}\left(\mathrm{r},t\right)\Psi^{'}\left(\mathrm{r},t\right)=i\hbar\frac{\partial\Psi^{'}\left(\mathrm{r},t\right)}{\partial t}
\end{equation}
 where {[}1{]} 
\begin{equation}
H^{'}\left(\mathrm{r},t\right)=RHR^{\dagger}+i\hbar\frac{\partial R}{\partial t}R^{\dagger}
\end{equation}
 Choosing 
\begin{equation}
R=\exp\left(-ie\chi\left(\mathrm{r},t\right)/\hbar\right)
\end{equation}
which amounts to choosing the Coulomb gauge, we have 
\begin{equation}
H^{'}=\frac{1}{2m}\left[\mathbf{P}+e\mathbf{A^{'}}\left(\mathrm{r},t\right)\right]^{2}-e\Phi^{'}\left(\mathrm{r},t\right)+V\left(r\right)
\end{equation}
 It is important to note that we will be working in the Coulomb gauge,
which is not relativistically covariant, but for which $\Phi\left(\mathrm{r},t\right)=0$
and $\nabla\cdot\mathbf{A}=0$ ( the transversality condition), therefore
the radiation field is completely determined by the vector potential
. If there are no sources near the atom, then $\mathbf{A}$ satisfies
the homogeneous wave equation 
\begin{equation}
\nabla^{2}\mathbf{A}-\frac{1}{c^{2}}\frac{\partial^{2}\mathbf{A}}{\partial t^{2}}=0.
\end{equation}
 whose solution has the form 
\begin{equation}
\mathbf{A}=\mathbf{A}_{0}e^{i\left(\mathbf{k\cdot r}-\omega_{k}t\right)}+\mathbf{A}_{0}^{\dagger}e^{-i\left(\mathbf{k\cdot r}-\omega_{k}t\right)}
\end{equation}
 In the Coulomb gauge the radiation field is completely determined
by the vector potential as can be seen from the Hamiltonian 
\begin{equation}
H\left(\mathrm{r},t\right)=\frac{\mathbf{P^{2}}}{2m}+\frac{e}{m}\mathbf{A}\cdot\mathbf{P}+\frac{e^{2}}{m}\mathbf{A}^{2}+V\left(r\right)
\end{equation}
 The transformed Hamiltonian becomes{[}1{]} 
\begin{equation}
H^{'}=\frac{1}{2m}\left[\mathbf{P}+e\left(\mathbf{A+\nabla\chi}\right)\right]^{2}-e\frac{\partial\chi}{\partial t}+V\left(r\right)
\end{equation}
 Since $\left|\mathbf{k}\right|=2\pi/\lambda$, it follows that for
$\left|\mathbf{r}\right|$$\sim5a_{0}$ ( Bohr radius) and $\lambda\thicksim500\mathrm{nm}$
(optical radiation) , $\mathbf{k\cdot r}\ll1$,Thus we may invoke
the dipole approximation which gives the first order interactions
and which also implies that on length scales $\sim\left|\mathbf{r}\right|_{atom}$,
the vector potential is locally uniform, and we may make use of the
fact that $\mathbf{A}\left(\mathrm{r},t\right)\simeq\mathbf{A}\left(t\right)$.
If we choose our gauge function to be 
\begin{equation}
\chi\left(\mathrm{r},t\right)=-\mathbf{A}\cdot\mathbf{r}
\end{equation}
 it follows that 
\begin{equation}
\nabla\chi\left(\mathrm{r},t\right)=-\mathbf{A}\left(t\right)
\end{equation}
\begin{equation}
\frac{\partial\chi}{\partial t}\left(\mathrm{r},t\right)=-\mathbf{r}\cdot\frac{\partial\mathbf{A}}{\partial t}=-\mathbf{r}\cdot\mathbf{E}\left(t\right)
\end{equation}
 which means that 
\begin{equation}
H^{'}\left(\mathrm{r},t\right)=\frac{\mathbf{P^{2}}}{2m}+V\left(r\right)+e\mathbf{r}\cdot\mathbf{E}\left(t\right)
\end{equation}
 we recognize the quantity $\mathbf{d}=-e\mathbf{r}$ as the electric
dipole moment.and we may write 
\begin{equation}
H^{'}\left(\mathrm{r},t\right)=H_{0}-\mathbf{d}\cdot\mathbf{E}\left(t\right)
\end{equation}

\end{doublespace}

\begin{doublespace}

\subsection{Interaction of an atom with a classical dipole field }
\end{doublespace}

\begin{doublespace}
Let us begin with the case of a classical field of frequency $\omega$,
given by {[}1{]} 
\begin{equation}
\mathbf{E}\left(t\right)=\mathbf{E}_{0}\cos\left(\omega t\right)\Theta\left(t\right)
\end{equation}
where 
\begin{equation}
\Theta\left(t\right)=\begin{cases}
1 & t>0\\
0 & t<0
\end{cases}
\end{equation}
just means that the field is turned on at a time $t=0.$ We can study
the interaction of an atom with this field by using perturbation theory.
Expanding to first order just amounts to using the dipole approximation,
$\mathbf{k\cdot r}\ll1,$ which we have seen previously is satisfied
in the case of atoms interacting with a classical electromagnetic
field. 

Given an atom, in some initial state $\left|i\right\rangle $, we
can expand of the atomic state of the atom for all $t>0,$in a basis
of uncoupled atomic states $\left|k\right\rangle $ , which span the
space of $H^{int}$,
\begin{equation}
\left|\psi\left(t\right)\right\rangle =\sum_{k}C_{k}\left(t\right)e^{-iE_{k}t/\hbar}\left|k\right\rangle 
\end{equation}
 where the amplitudes $C_{k}\left(t\right)$ are normalized such that
\begin{equation}
\sum_{k}\left|C_{k}\left(t\right)\right|^{2}=1.
\end{equation}

Now,working in the Schrodinger picture, the atomic state at time $t$
, $\left|\psi\left(t\right)\right\rangle $, must obey the time dependent
Schrodinger equation, which is:
\begin{equation}
i\hbar\frac{\partial\left|\psi\left(t\right)\right\rangle }{\partial t}=\left(H_{0}+H^{int}\right)\left|\psi\left(t\right)\right\rangle 
\end{equation}
 where, in the dipole approximation, as we know from the last section,
$H^{int}=-\mathbf{d}\cdot\mathbf{E}\left(t\right)$. Substituting
our expression for $\left|\psi\left(t\right)\right\rangle $ into
the Schrodinger equation, and then multiplying from the left by $\left\langle l\right|e^{-iE_{k}t/\hbar}$,we
( denoting time derivatives with dots $\:\dot{}\;$ ) obtain a set
of coupled first order differential equations for the amplitudes 
\[
\dot{C_{k}\left(t\right)}=-\frac{i}{\hbar}\sum_{k}C_{k}\left(t\right)\left\langle l\right|H^{int}\left|k\right\rangle e^{i\omega_{lk}t}
\]
 Where, 
\begin{equation}
\omega_{lk}=\left(E_{l}-E_{k}\right)/\hbar,
\end{equation}
are the transition frequencies between atomic states $\left|l\right\rangle $
and $\left|k\right\rangle .$ In order that we may solve these equations
we must also subject them to the condition, that the initial atomic
state is $\left|i\right\rangle ,$ which implies that $C_{k}\left(0\right)=1$.
As the state evolves in time, the initial state will transfer to other
other $\left|f\right\rangle $ with a probability given by 
\begin{equation}
P_{i\rightarrow f}\left(t\right)=\left|C_{f}\left(t\right)\right|^{2}.
\end{equation}

To further simplify the task of solving this set of coupled equations
analytically, we expand the amplitudes as a power series in some coupling
parameter $0<\lambda<1$(which measures the strength of the interaction
relative to scale at which our theory breaks down, $\lambda=1$).
\begin{equation}
C_{l}\left(t\right)=C_{l}^{\left(0\right)}\left(t\right)+\lambda C_{l}^{\left(1\right)}\left(t\right)+\lambda^{2}C_{l}^{\left(2\right)}\left(t\right)+...
\end{equation}

Inserting the expression for $C_{l}\left(t\right)$ into (IV.1) we
obtain a recursion for the $n^{th}$ amplitude 
\begin{equation}
\dot{C_{l}^{\left(n\right)}\left(t\right)}=-\frac{i}{\hbar}\sum_{k}C_{k}^{\left(n-1\right)}\left(t\right)\left\langle l\right|H_{lk}^{int}\left|k\right\rangle e^{i\omega_{lk}t}
\end{equation}
 which leads to a coupled set of equations for all of the $C_{l}^{n}\left(t\right)$,
which up to second order are given by 
\begin{equation}
\dot{C_{l}^{\left(0\right)}\left(t\right)}=0
\end{equation}
\begin{equation}
\dot{C}_{l}^{\left(1\right)}\left(t\right)=-\frac{i}{\hbar}\sum_{k}C_{k}^{\left(0\right)}\left(t\right)\left\langle l\right|H_{lk}^{int}\left|k\right\rangle e^{i\omega_{lk}t}
\end{equation}
\begin{equation}
\dot{C}_{l}^{\left(2\right)}\left(t\right)=-\frac{i}{\hbar}\sum_{k}C_{k}^{\left(1\right)}\left(t\right)\left\langle l\right|H_{lk}^{int}\left|k\right\rangle e^{i\omega_{lk}t}
\end{equation}
 The only surviving terms in the sum are those for $k=i$ . Therefore
the first order amplitude becomes, upon integrating on time 
\begin{equation}
C_{f}^{\left(1\right)}\left(t\right)=-\frac{i}{\hbar}\int_{0}^{t}dt'H_{fi}^{int}e^{i\omega_{fi}t'}C_{i}^{\left(0\right)}\left(t'\right)
\end{equation}
using the recursion relation, inserting this value for $C_{l}^{\left(1\right)}\left(t\right)$
into the equation for $\dot{C}_{l}^{\left(2\right)}\left(t\right)$
and integrating on time enables us to find $C_{l}^{\left(2\right)}\left(t\right)$,
(see {[}1{]})
\begin{equation}
C_{f}^{\left(2\right)}\left(t\right)=-\frac{i}{\hbar}\sum_{l}\int_{0}^{t}dt'H_{fl}^{int}\left(t'\right)e^{i\omega_{fl}t'}C_{l}^{\left(1\right)}\left(t'\right)
\end{equation}
\begin{equation}
=\left(-\frac{i}{\hbar}\right)^{2}\sum_{l}\int_{0}^{t}dt'\int_{0}^{t'}dt''H_{fl}^{int}\left(t'\right)e^{i\omega_{fl}t'}H_{li}^{int}\left(t''\right)e^{i\omega_{fl}t''}C_{l}^{\left(0\right)}\left(t''\right)
\end{equation}

The total transition probability as a function of time, for a transition
from state $\left|i\right\rangle $ to a state $\left|f\right\rangle $
is: 
\begin{equation}
P_{i\rightarrow f}\left(t\right)=\left|C_{f}^{\left(0\right)}\left(t\right)+C_{f}^{\left(1\right)}\left(t\right)+C_{f}^{\left(2\right)}\left(t\right)+...\right|^{2}
\end{equation}
 We have up to now, neglected taking account of the fact that the
dipole moment operator $\mathbf{d}$ only has non-vanishing matrix
elements for states of opposite parity. Taking this into account we
see that up to first order $C_{i}^{\left(0\right)}\left(t'\right)=1$,
so that 
\begin{equation}
C_{f}^{\left(1\right)}\left(t\right)=-\frac{i}{\hbar}\int_{0}^{t}\, dt'H_{f\, i}^{int}\, e^{i\omega_{fi}t'}
\end{equation}
\begin{equation}
=\frac{1}{2\hbar}\left(\mathbf{d}\cdot\mathbf{E}_{0}\right)_{fi}\left[\frac{\left(e^{i\left(\omega+\omega_{f}\right)t}-1\right)}{\left(\omega+\omega_{fi}\right)}-\frac{\left(e^{-i\left(\omega+\omega_{f}\right)t}-1\right)}{\left(\omega-\omega_{fi}\right)}\right]
\end{equation}
When the radiation frequency $\omega$ is near the atomic transition
frequency $\omega_{fi}$, we will have resonance, and in this case
we may neglect the first ``anti-resonant'' term since the second
term will clearly dominate. This is the so called \emph{rotating wave
approximation.} With this in mind, the first order transition probability
becomes {[}1{]} 
\begin{equation}
P_{i\rightarrow f}\left(t\right)=\frac{\left|\left(\left(\mathbf{d}\cdot\mathbf{E}_{0}\right)_{fi}\right)\right|}{\hbar^{2}}\frac{\sin^{2}\left(\Delta t/2\right)}{\Delta^{2}}
\end{equation}
 where we have introduced the notation $\Delta=\omega-\omega_{fi}$,
which is known as the \emph{detuning parameter. }
\end{doublespace}

\begin{doublespace}

\subsection{Interaction of an atom with a quantized dipole field }
\end{doublespace}

\begin{doublespace}
Earlier, while working in Heisenberg representation, we found that
the quanta of a single mode electric field, in the absence of sources
of any kind, were given by 
\begin{equation}
\mathbf{\hat{E}}\left(t\right)=i\left(\frac{\hbar\omega}{2\varepsilon_{0}V}\right)^{1/2}\mathbf{e}\left[ae^{-i\omega t}-a^{\dagger}e^{i\omega t}\right]
\end{equation}
switching to the Schrodinger representation, this becomes 
\begin{equation}
\mathbf{\hat{E}}=i\left(\frac{\hbar\omega}{2\varepsilon_{0}V}\right)^{1/2}\mathbf{e}\left[a-a^{\dagger}\right]
\end{equation}

The free Hamiltonian is 
\begin{equation}
H_{0}=H^{atom}+H^{field}
\end{equation}
where 
\begin{equation}
H^{atom}=\mathbf{P}^{2}/2m+V\left(r\right)
\end{equation}
 and 
\begin{equation}
H^{field}=\hbar\omega a^{\dagger}a
\end{equation}
are the source free Hamiltonians of the atomic system and the field,
respectively. We have suppressed the vacuum energy in our expression
for $H^{field}$ because it does not contribute to the dynamics. The
interaction Hamiltonian is 
\begin{equation}
H^{int}=-\mathbf{d}\cdot\mathbf{E}\left(t\right)=-i\left(\frac{\hbar\omega}{2\varepsilon_{0}V}\right)^{1/2}\left(\mathbf{d}\cdot\mathbf{e}\right)\left(a-a^{\dagger}\right)
\end{equation}
\begin{equation}
=\mathbf{d}\cdot\mathcal{E}_{0}\left(a^{\dagger}-a\right)
\end{equation}
where, 
\begin{equation}
\mathcal{E}_{0}=i\left(\hbar\omega/2\varepsilon_{0}V\right)^{1/2}\mathbf{e}
\end{equation}

We have thus quantized the atomic system as well as the field system.
If we wish to combine the distinct atom and field systems into one,
atom-field system, we must remember that the state space of the atom-field
system will in general be a linear superposition of the eigenstates
of $H_{atom}$ and $H_{field}$. Consider an atomic system in the
initial state $\left|a\right\rangle .$ If we combine this atomic
system with the field system which initially contains $n$ photons,
then we will have the atom-fields system which is initially in the
state 
\begin{equation}
\left|i\right\rangle =\left|a\right\rangle \left|n\right\rangle .
\end{equation}
 Since the interaction Hamiltonian $H^{int}$ is proportional to $\left(a^{\dagger}-a\right)$,
it follows that for the $n^{th}$ eigenstate $\left|n\right\rangle $,
the only non-vanishing matrix elements of $H^{int}$ (in the atom-field
eigenbasis) are the following 
\begin{equation}
\left\langle H^{int}\right\rangle =\sum_{i=1,2}\left\langle f_{i}\right|H_{int}\left|i\right\rangle =\left(\mathbf{d}\cdot\mathcal{E}_{0}\right)_{ba}\left\langle b\,,\, m\right|\left(a^{\dagger}-a\right)\left|a\,,\, n\right\rangle =
\end{equation}
\begin{equation}
=\left(\mathbf{d}\cdot\mathcal{E}_{0}\right)_{ba}\left(\sqrt{n+1}\delta_{n,.n+1}-\sqrt{n}\delta_{n,n-1}\right).
\end{equation}
where 
\begin{equation}
\left(\mathbf{d}\cdot\mathcal{E}_{0}\right)_{ba}=\left\langle b\right|\mathbf{d}\left|a\right\rangle \cdot\mathcal{E}_{0}
\end{equation}
 The quantity $\left\langle b\right|\mathbf{d}\left|a\right\rangle =\mathbf{d}_{ba}$
gives the transition dipole moments between the states $\left|b\right\rangle $
and $\left|a\right\rangle $.Therefore the interaction Hamiltonian
couples the $n^{th}$ state to either the $n+1$ or $n-1$ state.
In fact $H^{int}$ induces a transition from the initial state of
the atom-field system $\left|i\right\rangle $ to the state $\left|f_{1}\right\rangle =\left|b\right\rangle \left|n-1\right\rangle $
by absorption of a photon or to the state $\left|f_{2}\right\rangle =\left|b\right\rangle \left|n+1\right\rangle $,
by the emission of a photon. The energies of these states are {[}1{]}
\begin{equation}
\left|i\right\rangle =\left|a\right\rangle \left|n\right\rangle \quad\leftrightarrow\quad E_{i}=E_{a}+n\hbar\omega
\end{equation}
\begin{equation}
\left|f_{1}\right\rangle =\left|b\right\rangle \left|n-1\right\rangle \quad\leftrightarrow\quad E_{f_{1}}=E_{b}+\left(n-1\right)\hbar\omega
\end{equation}
\begin{equation}
\left|f_{2}\right\rangle =\left|b\right\rangle \left|n+1\right\rangle \quad\leftrightarrow\quad E_{f_{1}}=E_{b}+\left(n+1\right)\hbar\omega
\end{equation}
 where $E_{a}$ and $E_{b}$ are the energy eigenvalues of the respective
atomic states $\left|a\right\rangle $ and $\left|b\right\rangle $.

Let us compare the results of the semi-classical versus the quantum
treatment of this problem. In both cases, absorption is forbidden
in any state for which $n=0$ (zero photons in the system), for obvious
reasons. However, for the quantum case of emission, even if $n=0$
transitions may occur known as \emph{spontaneous emission} a phenomenon
with no semi-classical analog. In cases where $n>0,$ we then speak
of the \emph{stimulated} \emph{emission} of an additional photon.
In the classical case, if you start with no field, i.e no photon,
then you will never have a photon later, but a photon later is almost
a certainty in the quantum case. The matrix elements of the interaction
are in the case of adsorption {[}1{]} 
\begin{equation}
\left\langle f_{1}\right|H^{int}\left|i\right\rangle =\left\langle b,n-1\right|H^{int}\left|a,n\right\rangle 
\end{equation}
\begin{equation}
=-\left(\mathbf{d}\cdot\mathcal{E}_{0}\right)_{ba}\sqrt{n}
\end{equation}
and for the case of emission, are 
\begin{equation}
\left\langle f_{2}\right|H^{int}\left|i\right\rangle =\left\langle b,n+1\right|H^{int}\left|a,n\right\rangle 
\end{equation}
\begin{equation}
=\left(\mathbf{d}\cdot\mathcal{E}_{0}\right)_{ba}\sqrt{n+1}
\end{equation}
where, just as before, 
\begin{equation}
\left(\mathbf{d}\cdot\mathcal{E}_{0}\right)_{ba}=\left\langle b\right|\mathbf{d}\left|a\right\rangle \cdot\mathcal{E}_{0}
\end{equation}

Fermi's golden rule tells us that the rates of emission and absorption
are proportional to square modulus of the matrix element coupling
initial $\left|i\right\rangle $ and final states $\left|f_{1}\right\rangle ,\left|f_{2}\right\rangle ,$
which in the case of a single mode (\emph{monochromatic)} field coupled
to an atom, whose final state space is spanned by $\left|a\right\rangle $,
$\left|b\right\rangle $, The transition matrix elements are given
by: (see {[}1{]}) 
\begin{equation}
W_{i\rightarrow\left[f\right]}=\frac{\pi}{2}\sum_{\left[f\right]}\frac{\left|\left(\mathbf{d}\cdot\mathcal{E}_{0}\right)_{fi}\right|^{2}}{\hbar^{2}}\delta\left(\omega-\omega_{fi}\right).
\end{equation}
Moreover, since, 
\begin{equation}
\left(\mathbf{d}\cdot\mathcal{E}_{0}\right)_{ba}=\left\langle b\right|\mathbf{d}\left|a\right\rangle \cdot\mathcal{E}_{0}
\end{equation}
this becomes, 
\begin{equation}
W_{i\rightarrow\left[f\right]}=\frac{\pi}{2}\sum_{\left[f\right]}\frac{\left|\left\langle b\right|\mathbf{d}\left|a\right\rangle \cdot\mathcal{E}_{0}\right|^{2}}{\hbar^{2}}\delta\left(\omega-\omega_{fi}\right).
\end{equation}
 Therefore the ratio of these rates is given by 
\begin{equation}
\frac{\left|\left\langle f_{2}\right|H^{int}\left|i\right\rangle \right|^{2}}{\left|\left\langle f_{1}\right|H^{int}\left|i\right\rangle \right|^{2}}=\frac{n+1}{n}.
\end{equation}

For the time being, let's just focus on two atomic states $\left|a\right\rangle $
and $\left|b\right\rangle $, and ignore the rest. This allows us
to write the write the state vector as 
\begin{alignat}{1}
\left|\psi\left(t\right)\right\rangle  & =C_{i}\left(t\right)\left|a\right\rangle \left|n\right\rangle e^{-iE_{a}t/\hbar}e^{-in\omega t}+C_{f1}\left(t\right)\left|b\right\rangle \left|n-1\right\rangle e^{-iE_{b}t/\hbar}e^{-i\left(n-1\right)\omega t}\nonumber \\
 & \,+C_{f2}\left(t\right)\left|b\right\rangle \left|n+1\right\rangle e^{-iE_{b}t/\hbar}e^{-i\left(n+1\right)\omega t}
\end{alignat}
 since we already said that the initial state was $\left|\psi\left(t\right)\right\rangle =\left|a\right\rangle \left|n\right\rangle $,
so therefore we have: 
\begin{equation}
C_{i}\left(0\right)=1
\end{equation}
\begin{equation}
C_{f1}\left(0\right)=C_{f2}\left(0\right)=0
\end{equation}
We can use perturbation theory just before to obtain the probability
amplitudes for all times, $t>0$ . 
\begin{equation}
C_{f1}^{\left(1\right)}\left(t\right)=-\frac{i}{\hbar}\int_{0}^{t}dt'\left\langle f_{1}\right|H^{int}\left|i\right\rangle e^{i\omega_{f1i}t'}C_{i}^{\left(0\right)}\left(t'\right)\qquad\mathrm{\left(absorption\right)}
\end{equation}
\begin{equation}
C_{f2}^{\left(1\right)}\left(t\right)=-\frac{i}{\hbar}\int_{0}^{t}dt'\left\langle f_{2}\right|H^{int}\left|i\right\rangle e^{i\omega_{f2i}t'}C_{i}^{\left(0\right)}\left(t'\right)\qquad\mathrm{\left(emission\right)}
\end{equation}
where $\omega_{f1i}=\left(E_{f1}-E_{i}\right)/\hbar$ and $\omega_{f2i}=\left(E_{f2}-E_{i}\right)/\hbar$
. Therefore the probability that the atom under goes a transition
to the final state $\left|b\right\rangle $ is the sum $C_{f}^{\left(1\right)}=C_{f1}^{\left(1\right)}+C_{f2}^{\left(1\right)}$
, i.e. 
\begin{equation}
C_{f}^{\left(1\right)}==\frac{i}{\hbar}\left(\mathbf{d}\cdot\mathbf{E}_{0}\right)_{ab}\left[\sqrt{n+1}\frac{\left(e^{i\left(\omega+\omega_{ba}\right)t}-1\right)}{\left(\omega+\omega_{ba}\right)}-\sqrt{n}\frac{\left(e^{-i\left(\omega+\omega_{ba}\right)t}-1\right)}{\left(\omega-\omega_{ba}\right)}\right]
\end{equation}
 where $\omega_{ba}=\left(E_{b}-E_{a}\right)/\hbar$. If the initial
state $\left|a\right\rangle $ happens to be the excited state then
$\omega_{ba}<0$. In this case, for radiation frequencies $\omega\sim\left(-\omega_{ba}\right)$
, and we get spontaneous emission. 
\end{doublespace}

\begin{doublespace}

\subsection{The Jaynes-Cummings Model}
\end{doublespace}

\begin{doublespace}
Perturbation theory breaks down in situation where we have a driving
field of near resonance frequency. This is because the resonance causes
large population transfers and the assumption made in perturbation
theory that $C_{i}\left(t\right)\approx1$, no longer holds. Therefore
we must take another approach. One such approach can be understood
by noticing that for the case of near resonance, most of the population
is transferred to the near resonant state, so that the other states
may be neglected. The two most dominant states remain, and the resulting
system of differential equations are much more tractable.

The \emph{Jaynes-Cummings} Hamiltonian is (see{[}1{]},{[}10{]})
\begin{equation}
H=\frac{1}{2}\hbar\omega_{0}\sigma_{3}+\hbar\omega a^{\dagger}a+\hbar\lambda\left(\sigma_{+}a+\sigma_{+}a^{\dagger}\right).
\end{equation}

The \emph{interaction term }is 
\begin{equation}
H^{int}=\hbar\lambda\left(\sigma_{+}a+\sigma_{+}a^{\dagger}\right)
\end{equation}
and it induces the transitions, 
\begin{equation}
\left|\, e\,\right\rangle \left|\, n\,\right\rangle \leftrightarrow\left|\, g\,\right\rangle \left|\, n+1\,\right\rangle 
\end{equation}
 or 
\begin{equation}
\left|\, e\,\right\rangle \left|\, n-1\,\right\rangle \leftrightarrow\left|\, g\,\right\rangle \left|\, n\,\right\rangle .
\end{equation}
 The product states $\left|\, e\,\right\rangle \left|\, n\,\right\rangle ,\;\left|\, g\,\right\rangle \left|\, n+1\,\right\rangle ,$etc.
, span what is known the \emph{bare }basis, they are \emph{bare }states
of the Jaynes-Cummings model. For any given $n$, the dynamics of
the system is confined to a $2D$ space of product states spanned
by 
\begin{equation}
\left\{ \left|\, e\,\right\rangle \left|n-1\right\rangle ,\left|\, g\,\right\rangle \left|n-1\right\rangle ,\left|\, e\,\right\rangle \left|\, n\,\right\rangle ,\left|\, g\,\right\rangle \left|\, n\,\right\rangle \right\} .
\end{equation}
We can therefore define general product states for any $n$ 
\begin{equation}
\left|\psi_{1n}\right\rangle =\left|\, e\,\right\rangle \left|\, n\,\right\rangle 
\end{equation}
\begin{equation}
\left|\psi_{2n}\right\rangle =\left|\, g\,\right\rangle \left|n+1\right\rangle .
\end{equation}
It follows that 
\begin{equation}
\left\langle \psi_{1n}\right|\left|\psi_{2n}\right\rangle =0
\end{equation}
The matrix representation of $H$ 
\begin{equation}
H_{ij}=\left\langle \psi_{in}\right|H\left|\psi_{jn}\right\rangle 
\end{equation}
In this basis becomes 
\begin{equation}
H_{ij}=\left[\begin{array}{cc}
n\omega+\frac{1}{2}\hbar\omega_{0} & \hbar\lambda\sqrt{n+1}\\
\hbar\lambda\sqrt{n+1} & \left(n+1\right)\omega-\frac{1}{2}\hbar\omega_{0}
\end{array}\right]
\end{equation}
For any given $n$\emph{, }the energy eigenvalues are 
\begin{equation}
E_{\pm}\left(n\right)=\left(n+\frac{1}{2}\right)\hbar\omega\pm\hbar\Omega_{n}\left(\Delta\right)
\end{equation}
where $\Delta=\left(\omega_{0}-\omega\right)$, is the \emph{detuning
parameter }of the atomic transition frequency and the monochromatic
field and 
\begin{equation}
\Omega_{n}\left(\Delta\right)=\left[\Delta^{2}+4\lambda^{2}\left(n+1\right)\right]^{1/2}
\end{equation}
 is the damped \emph{Rabi oscillation frequency}, which in the case
of resonance, i.e, $\Delta=0$ , becomes 
\begin{equation}
\Omega_{n}\left(0\right)=2\lambda\sqrt{n+1}
\end{equation}
The set of energy eigenstates form what are known as the \emph{dressed
states, }and\emph{ }these are a linear combination of the \emph{bare
states} which are 
\begin{equation}
\left|n,+\right\rangle =\cos\left(\Phi_{n}/2\right)\left|\psi_{1n}\right\rangle +\sin\left(\Phi_{n}/2\right)\left|\psi_{2n}\right\rangle 
\end{equation}
\begin{equation}
\left|n,-\right\rangle =-\sin\left(\Phi_{n}/2\right)\left|\psi_{1n}\right\rangle +\cos\left(\Phi_{n}/2\right)\left|\psi_{2n}\right\rangle 
\end{equation}
 with $\Phi_{n}$ given by 
\begin{equation}
\Phi_{n}=\tan^{-1}\left(\frac{2\lambda\sqrt{n+1}}{\Delta}\right)=\tan^{-1}\left(\frac{\Omega_{n}\left(0\right)}{\Delta}\right)
\end{equation}
Moreover,
\begin{equation}
\sin\left(\Phi_{n}/2\right)=\frac{1}{\sqrt{2}}\left[\frac{\Omega_{n}\left(\Delta\right)-\Delta}{\Omega_{n}\left(\Delta\right)}\right]^{1/2}
\end{equation}
\begin{equation}
\cos\left(\Phi_{n}/2\right)=\frac{1}{\sqrt{2}}\left[\frac{\Omega_{n}\left(\Delta\right)+\Delta}{\Omega_{n}\left(\Delta\right)}\right]^{1/2}
\end{equation}
The dressed states $\left|n,\pm\right\rangle $ comprise the Jaynes-Cummings
doublet. The $\hbar\Omega_{n}\left(\Delta\right)$ term splits the
energies of the bare states$\left|\;\psi_{1n}\right\rangle ,\;\left|\psi_{2n}\right\rangle ,\;$an
effect known as the \emph{dynamic stark shift}. In the case of exact
resonance $\Delta=0,\;$ the bare states become degenerate, but the
dynamic stark shift splitting of the dressed states endures. In the
exact resonance limit the dressed states can be represented in the
basis of bare states as
\begin{equation}
\left|n,+\right\rangle =\frac{1}{\sqrt{2}}\left(\left|e\right\rangle \left|n\right\rangle +\left|g\right\rangle \left|n+1\right\rangle \right)
\end{equation}
\begin{equation}
\left|n,-\right\rangle =\frac{1}{\sqrt{2}}\left(-\left|e\right\rangle \left|n\right\rangle +\left|g\right\rangle \left|n+1\right\rangle \right)
\end{equation}
 To obtain the dynamics in the dressed state basis, let us consider
the case of an atom field system, for which the field is prepared
in some superposition of initial states as 
\begin{equation}
\left|\psi_{f}\left(0\right)\right\rangle =\sum_{n}C_{n}\left|\, n\,\right\rangle 
\end{equation}
 and for which an atom, initially in the state $\left|e\right\rangle $
gets injected into the field. Therefore the initial state of the atom
field system is 
\begin{equation}
\left|\psi_{a\, f}\left(0\right)\right\rangle =\left|\psi_{f}\left(0\right)\right\rangle \left|\, e\,\right\rangle 
\end{equation}
\begin{equation}
=\sum_{n}C_{n}\left|n\right\rangle \left|e\right\rangle =\sum_{n}C_{n}\left|\psi_{1n}\right\rangle .
\end{equation}
 Now, from (IV.2) and (IV.3), it follows that 
\begin{equation}
\left|\psi_{1n}\right\rangle =\cos\left(\Phi_{n}/2\right)\left|n,+\right\rangle -\sin\left(\Phi_{n}/2\right)\left|n,-\right\rangle 
\end{equation}
 therefore the initial state of the atom-field system is 
\begin{equation}
\left|\psi_{af}\left(0\right)\right\rangle =\sum_{n}C_{n}\left[\cos\left(\Phi_{n}/2\right)\left|n,+\right\rangle -\sin\left(\Phi_{n}/2\right)\left|n,-\right\rangle \right]
\end{equation}
One nice feature of the dressed states, is that they are stationary
states of the Hamiltonian, and a consequence, the time evolution of
\emph{H }is 
\begin{equation}
\left|\psi_{af}\left(t\right)\right\rangle =\exp\left(-\frac{i}{\hbar}Ht\right)\left|\psi_{af}\left(0\right)\right\rangle 
\end{equation}
\begin{equation}
=\sum_{n}C_{n}\left[\cos\left(\Phi_{n}/2\right)\left|n,+\right\rangle e^{-iE_{+}\left(n\right)t/\hbar}-\sin\left(\Phi_{n}/2\right)\left|n,-\right\rangle e^{-iE_{-}\left(n\right)t/\hbar}\right]
\end{equation}

\end{doublespace}

\begin{doublespace}

\subsection{Jaynes-Cummings with large de-tuning }
\end{doublespace}

\begin{doublespace}
In the forgoing we have been working with an ``on resonance'' approximation,
in which the detuning parameter vanishes i.e., when $\Delta=0.$ A
more interesting case is the one in which detuning exists to the extent
that direct atomic transitions do not occur, but where \emph{dispersive}
interactions between a single and a cavity field do occur . 
\end{doublespace}

\begin{doublespace}
The effective Hamiltonian in the case of large detuning is given by
\begin{equation}
H_{eff}=\hbar\chi\left[\sigma_{+}\sigma_{-}+a^{\dagger}a\sigma_{3}\right]
\end{equation}
Where, 
\begin{equation}
\chi=\lambda^{2}/\Delta
\end{equation}

The transition operators are the projections 
\begin{equation}
\sigma_{+}=\left|e\right\rangle \left\langle g\right|\qquad\sigma_{-}=\left|g\right\rangle \left\langle e\right|
\end{equation}
 Note that 
\begin{equation}
\sigma_{+}\sigma_{-}=\left|e\right\rangle \left\langle e\right|
\end{equation}
 is the emission projector, and the inversion operator $\sigma_{3}$
is 
\begin{equation}
\sigma_{3}=\left|e\right\rangle \left\langle e\right|-\left|g\right\rangle \left\langle g\right|
\end{equation}

The transition and inversion operators obey the Pauli algebra
\begin{equation}
\left[\sigma_{+},\sigma_{-}\right]=\sigma_{3}
\end{equation}
\begin{equation}
\left[\sigma_{3},\sigma_{\pm}\right]=2\sigma_{\pm}
\end{equation}

Suppose that the initial state of the atom-field system has the configuration
of an atom in the ground state and the field in a number state, i.e.
\begin{equation}
\left|\psi\left(0\right)\right\rangle =\left|g\right\rangle \left|n\right\rangle 
\end{equation}
The time evolved state becomes
\begin{equation}
\left|\psi\left(t\right)\right\rangle =e^{-iH_{eff}t/\hbar}\left|\psi\left(0\right)\right\rangle =e^{i\chi nt}\left|g\right\rangle \left|n\right\rangle .
\end{equation}
 Similarly, for the initial conditions 
\begin{equation}
\left|\psi\left(0\right)\right\rangle =\left|e\right\rangle \left|n\right\rangle 
\end{equation}
We get 
\begin{equation}
\left|\psi\left(t\right)\right\rangle =e^{-iH_{eff}t/\hbar}\left|\psi\left(0\right)\right\rangle =e^{i\chi\left(n+1\right)t}\left|e\right\rangle \left|n\right\rangle .
\end{equation}
and nothing happens except the production of unmeasurable phase factors. 

However, if the initial state is a \emph{coherent state} of the field,
that is in the case where 
\begin{equation}
\left|\psi\left(0\right)\right\rangle =\left|g\right\rangle \left|\alpha\right\rangle \quad\mathrm{Coherent\, initial\: state}
\end{equation}
We obtain 
\begin{equation}
\left|\psi\left(t\right)\right\rangle =e^{-iH_{eff}t/\hbar}\left|\psi\left(0\right)\right\rangle =\left|g\right\rangle \left|\alpha e^{i\chi t}\right\rangle 
\end{equation}

Similarly, for the initial state 
\begin{equation}
\left|\psi\left(0\right)\right\rangle =\left|e\right\rangle \left|\alpha\right\rangle 
\end{equation}
 we have 
\begin{equation}
\left|\psi\left(t\right)\right\rangle =e^{-iH_{eff}t/\hbar}\left|\psi\left(0\right)\right\rangle =e^{-i\chi t}\left|e\right\rangle \left|\alpha e^{-i\chi t}\right\rangle 
\end{equation}
 For either case of the initial coherent field state, the coherent
state amplitude gets rotated in phase space by the angle $\theta=\chi t$.
The direction of rotation depends on which initial states the atom
is in. 

Let us now consider the case of an atom in an initial superposition
of ground and excited states, which for simplicity, we assume takes
the form of a \emph{balanced} state with the form:
\begin{equation}
\left|\psi_{atom}\right\rangle =\frac{1}{\sqrt{2}}\left(\left|g\right\rangle +e^{i\phi}\left|e\right\rangle \right)\qquad\phi\leftrightarrow\mathrm{phase}
\end{equation}
 For an initial state 
\begin{equation}
\left|\psi\left(0\right)\right\rangle =\left|\psi_{atom}\right\rangle \left|\alpha\right\rangle 
\end{equation}
We obtain 
\begin{equation}
\left|\psi\left(t\right)\right\rangle =e^{-iH_{eff}t/\hbar}\left|\psi\left(0\right)\right\rangle =\frac{1}{\sqrt{2}}\left(\left|g\right\rangle \left|\alpha e^{i\chi t}\right\rangle +e^{-i\left(\chi t-\phi\right)}\left|e\right\rangle \left|\alpha e^{-i\chi t}\right\rangle \right)
\end{equation}
which is a much more interesting state, since now the atom and field
are \emph{entangled}. 

Taking $\chi t=\pi/2$ , for which $e^{i\chi t}=i$ , and $e^{-i\chi t}=-i$
, we have the $entangled$ state 
\begin{equation}
\left|\psi\left(\frac{\pi}{2\chi}\right)\right\rangle ==\frac{1}{\sqrt{2}}\left(\left|g\right\rangle \left|i\alpha\right\rangle -ie^{i\phi}\left|e\right\rangle \left|-i\alpha\right\rangle \right)
\end{equation}
which can be understood by analogy to the Schrodinger's cat paradox.
With this analogy, our atomic states correspond to the radioactive
atom in the paradox, and the two phase-separated coherent field states
play the role of Schrodinger's cat.Moreover the above entangled state
corresponds the entangled state 
\begin{equation}
\left|\psi_{\mathrm{atom-cat}}\right\rangle =\frac{1}{\sqrt{2}}\left(\left|\mathrm{atom\; not\; decayed}\right\rangle \left|\mathrm{cat\; alive}\right\rangle +\left|\mathrm{atom\; decayed}\right\rangle \left|\mathrm{cat\; dead}\right\rangle \right)
\end{equation}
 Coherent states differing in phase by $\pi$ are maximally distinguishable,
and there is effectively no overlap between the states, that is for
$\left|\alpha\right|$ sufficiently large. Very large values of $\left|\alpha\right|$
are \emph{macroscopically distinguishable}, while moderate values
of $\left|\alpha\right|$, \emph{mesoscopically distinguishable}. 
\end{doublespace}

\begin{doublespace}

\section{Application of CQED to Quantum Information Processing}
\end{doublespace}

\begin{doublespace}

\subsection{The Fabry-Perot Cavity}
\end{doublespace}

\begin{doublespace}
We start this section with a brief overview the Fabry-Perot cavity.
One of the most essential components of a Fabry-Perot cavity is a
partially silvered mirror, which partially reflects and transmits
incidents light $E_{a}$ and $E_{b}$, which has the effect of producing
output fields $E_{a'}$ and $E_{b'}$, which are related by the unitary
transformation: 
\begin{equation}
\left[\begin{array}{c}
E_{a'}\\
E_{b'}
\end{array}\right]=\left[\begin{array}{cc}
\sqrt{R} & \sqrt{1-R}\\
\sqrt{1-R} & -\sqrt{R}
\end{array}\right]\left[\begin{array}{c}
E_{a}\\
E_{b}
\end{array}\right]
\end{equation}
 where $R$ is the reflectivity of the mirror. 

A Fabry-Perot (FP) cavity is made from two plane parallel mirrors
of reactivities $R_{1}$ and $R_{2}$, incident upon which is light
form outside the cavity $E_{int}$. Inside the cavity, light bounces
back and forth between the two mirrors acquiring a phase shift $e^{i\phi}$
on each trip. The internal cavity field is 
\begin{equation}
E_{cav}=\sum_{k}E_{k}=\frac{\sqrt{1-R}E_{in}}{1+e^{i\phi}\sqrt{R_{1}R_{2}}}
\end{equation}
 One of the most important things about the Fabry Perot cavity for
purposes of CQED is the power in the internal cavity field mode as
a function of of the power and frequency of the input field, 
\begin{equation}
\frac{P_{cav}}{P_{in}}=\left|\frac{E_{cav}}{E_{in}}\right|^{2}=\frac{1-R_{1}}{\left|1+e^{i\phi}\sqrt{R_{1}R_{2}}\right|^{2}}
\end{equation}

Frequency selectivity arises because of constructive and destructive
interference between the cavity mode and the reflected light front.
Another indispensable feature is that on resonance, the cavity field
achieves a maximum which is approximately $\left(1-R\right)^{-1}$
times the incident field . 
\end{doublespace}

\begin{doublespace}

\subsection{Quantum Computation }
\end{doublespace}

\begin{doublespace}
Quantum information can be encoded with single photons in the \emph{dual
rail representation} 
\begin{equation}
c_{0}\left|01\right\rangle +c_{1}\left|10\right\rangle 
\end{equation}
Arbitrary unitary transformations can be applied to such quantum information
using phase shifters, beam splitters, and nonlinear optical Kerr media
{[}1{]}. 

The \emph{single photon representation} of a qubit is attractive because
it represents the information saturation limit of the electromagnetic
field and single photons, by today's standards can be generated relatively
easily and moreover and most importantly, \textbf{arbitrary qubit
operations become possible, in general in the dual-rail representation}.
The difficult part in this approach is making the photon - photon
scattering amplitudes large enough for entanglement to occur. In an
optical cavity this is commonly implemented with optical nonlinear
Kerr media. However, in reality even the best non-linear Kerr media
are weak, and are unable to provide a cross phase modulation of $180^{\circ}$
between single photon states. It is estimated , that even in the best
cases, approximately 50 photons would have to be absorbed for each
$180^{\circ}$ cross phase modulated photon. 

Despite its drawbacks, the optical quantum computer does provide us
with some insight into the architecture and design of a quantum computer.
Assuming we had sufficiently good components available, we could construct
an optical quantum computer, which will be almost entirely comprised
of optical interferometers. The quantum information is encoded in
both the photon number states and the photon phase. Interferometers
perform the function of switching between the two representations.
Stability however, becomes a major issue, and if a massive representation
of a qubit is chosen, then stable interferometers would be a challenge
to construct because of the relatively short scale of the de Broglie
wavelengths of the qubits.

\end{doublespace}


\begin{thebibliography}{10}
\begin{doublespace}
\bibitem[1]{key-1}Gerry, C.G. and Knight, P.L. \emph{Introductory
Quantum Optics, }Cambridge University Press (2005).

\bibitem[2]{key-2}Cohen-Tannoudji, Claude. \emph{Quantum Mechanics,
}Vol. 1, (New York: Wiley Interscience, 1977).

\bibitem[3]{key-3}P. A. M. Dirac, Proc. R. Soc. Lond., A 114 (1927),
243.

\bibitem[4]{key-4}L. Susskind and J. Glogower, Physics, 1 (1964),
49.

\bibitem[5]{key-5}D. T. Pegg and S. M. Barnett, Europhys. Lett.,
6 (1988), 483; Phys. Rev. A, 39 (1989), 1665; 43 (1991), 2579.

\bibitem[6]{key-6}J. H. Shapiro and S. R. Shepard, Phys. Rev. A,
43 (1991), 3795.

\bibitem[7]{key-7}C. W. Helstrom, Quantum Detection and Estimation
Theory (New York: Academic Press, 1976).

\bibitem[8]{key-8}Nielsen M. A., Chuang, I., \emph{Quantum Computation
and Quantum Information, }Cambridge University Press (2000).

\bibitem[10]{key-10}Griffiths, R.B. ,\emph{ Consistent Quantum Theory,
}Cambridge University Press (2002)\end{doublespace}
\end{thebibliography}
\end{document}